\documentstyle{pde96}
\Year{1996}
 \newcommand{\la}{\langle}
 \newcommand{\ra}{\rangle}
 \newcommand{\beq}{\begin{equation}}
 \newcommand{\eeq}{\end{equation}}
 \newcommand{\beqa}{\begin{eqnarray}}
 \newcommand{\eeqa}{\end{eqnarray}}
 \newcommand{\rf}[1]{(\ref{#1})}
 \newcommand{\D}[1]{{\cal D}(#1)}
 \newcommand{\M}{{\cal M}}
 \newcommand{\Dp}[1]{{\cal D}'(#1)}
 \newcommand{\lt}{\Lambda^{(2)}}
 \newcommand{\ls}{\Lambda^{(2)}_\Sigma}
 \newcommand{\ol}{\overline}
\begin{document}
\Title{Application of Microlocal Analysis to the Theory of Quantum
  Fields Interacting with a Gravitational Field\footnote{Talk
presented at the Conference on Partial Differential Equations, Potsdam
1996, to be publ. in the Proceedings}}
\Shorttitle{Microlocal Analysis and QFT on Curved Spacetimes}   
\By{{\sc Wolfgang Junker}}
\Names{W. Junker} 
\Email{wjunker@x4u.desy.de}
\maketitle
%
%

\begin{abstract}
It is explained how techniques from microlocal analysis can be used to
settle some long-standing questions that arise in the study of the
interaction of quantum matter fields with a classical 
gravitational background field.
\end{abstract}

\newsection{Introduction}
Quantum field theory (QFT) is the theory of elementary particles and
their fundamental interactions.  The very successful standard model
which describes the electromagnetic, weak and strong interactions of
the observed elementary particles does however not incorporate gravity.
The quantisation of gravity (``quantum gravity'') is a very difficult
problem which is presently still far from its solution. An easier, but
only approximate approach to the interaction of matter and
gravitational fields is the ``semiclassical theory'' (or ``QFT in
curved spacetimes'') where the gravitational field is described {\it
  classically} as a 4-dim. Lorentzian manifold $({\cal M},g)$ and only
the matter fields are {\it quantized} to operator valued Wightman
fields. This theory, which we will describe in the following more
closely, should have a wide range of physical applicability, from
quantum effects in the early universe to the Hawking radiation of
massive stars collapsing to a black hole, and should also lead to a
better understanding of the local features of standard QFT (for a good
introduction see e.g.~\cite{Fulling89}).

\newsection{The basic setting of QFT}
We only deal with the simple example of a free (= linear) massive
scalar field (Klein-Gordon field).\\
Let $({\cal M},g)$ be a 4-dim. globally hyperbolic (i.e.~there exists
a spacelike Cauchy hypersurface $\Sigma$ such that
$\M=\R\times\Sigma$) 
Lorentzian manifold ($g_{
\mu\nu}$ having signature $+---$) and consider the linear Klein-Gordon equation
\begin{eqnarray}
(\Box_g+m^2)\Phi &=& \left(
g^{\mu\nu}\nabla_\mu\nabla_\nu+m^2\right)\Phi\nonumber\\
&=& \frac{1}{\sqrt{g}}\partial_\mu\left( \sqrt{g}
g^{\mu\nu}\partial_\nu\Phi\right) +m^2\Phi = 0 \label{2.1}
\end{eqnarray}
in some coordinate system, where $m>0$ is the mass of the scalar field
$\Phi:{\cal M} \to {\R}$, $\nabla_\mu$ the covariant derivative on
${\cal M}$ w.r.t.~$g_{\mu\nu}$ and $\sqrt{g}:=|\det
(g_{\mu\nu})|^{1/2}$. 
By the
linearity of the field equation (\ref{2.1}) the field $\Phi$ has no
self-interaction (``free'' field), but is coupled via the metric
tensor to the gravitational field. (\ref{2.1}) is a hyperbolic partial
differential equation with variable coefficients. It possesses unique
retarded and advanced fundamental solutions $E^{ret}, E^{av}$.\\
Quantizing the classical field $\Phi$ means to construct a Hilbert
space ${\cal H}$ of physical states and an operator-valued
distribution $\hat{\Phi}(f)$ ($f\in {\cal D}({\cal M})$ a
testfunction) acting on $\cal H$ which describes the field observables
localized in supp $f$ $\subset {\cal M}$ subject to the following
axioms:\\
(i) $\forall f\in {\cal D}({\cal M}): \hat{\Phi}(f)$ is a linear
(unbounded) closable operator on ${\cal H}$ with dense domain
$D\subset {\cal H}$ such
that $\hat{\Phi}(f)^* \supset \hat{\Phi}(\bar{f})$ (hermiticity) and
$\hat{\Phi}(f) D\subset D$.\\
(ii) $\forall \psi\in D: \Lambda
^{(n)}(f_1,\ldots,f_n):=\left\langle\psi,
\hat{\Phi}(f_1)\ldots\hat{\Phi}(f_n)\psi\right\rangle
\in {\cal D}'({\cal M}^n)$ (n-point distributions).\\
(iii) $\forall f_1,f_2\in {\cal D}({\cal M}):
\left[\hat{\Phi}(f_1),\hat{\Phi}(f_2)\right] =
\frac{1}{2}\left\langle f_1, E f_2\right\rangle$, $E:=E^{ret}-E^{av}$
 (commutation relations).\\
(iv) $\forall f\in {\cal D}({\cal M}): \hat{\Phi}\left( (\Box_g+m^2)
f\right)=0$ (field equations).\\
(v) If ${\cal M}={\R}^4$ (Minkowski space) with metric $g=
\mbox{diag}(+1,-1,-1,-1)$ one demands Poincar\'{e} covariance of the
theory, in particular that the translations $T_a:x \mapsto x+a, a\in
{\R}^4,$ can be implemented in $\cal H$ by a strongly continuous
unitary group $U(a)=exp(i a_\mu P^\mu)$ whose generator $P^\mu$ has
spectrum in the positive forward light cone ({\it spectrum
  condition}). The {\it vacuum state} $\Omega \in D\subset {\cal H}$ is the
unique eigenstate of $P^\mu$ to eigenvalue 0. \\
Whereas conditions (i)--(iv) can equally well be formulated on a curved
manifold $\cal M$ as on ${\R}^4$ this axiom (v) depends in an
essential way on the special global structure of ${\R}^4$. Thus, the 
vacuum state does not exist on a generic manifold, and it has
been the main problem of QFT in curved spacetimes to find a substitute
for (v) in this case.\\
To study this question we consider only {\it quasifree} states $\psi
  \in D$ where (by def.) all the physical information is
  contained in the {\it 2-point correlation function}
  $\Lambda^{(2)}\in {\cal D}'({\cal M\times M})$.

\newsection{Hadamard states and their wavefront set}
The background geometry reacts on the energy-momentum content of the
matter fields via the semiclassical Einstein equations
\begin{equation}
R_{\mu\nu}-\frac{1}{2}g_{\mu\nu}R= 8\pi
\la\psi|\hat{T}_{\mu\nu}(x)|\psi\ra \label{3.1}
\eeq
where $\la\psi |\hat{T}_{\mu\nu}(x) |\psi\ra$ is the expectation value
in some state $\psi\in {\cal H}$ of the energy-momentum tensor
operator $\hat{T}_{\mu\nu}$ of $\hat{\Phi}$ at a point $x\in \M$, the
l.h.s.~of \rf{3.1} is the Einstein tensor. Looking at the classical
expression for $T_{\mu\nu}$
\[
T_{\mu\nu}(x)=(\nabla_\mu\Phi)(\nabla_\nu\Phi)-\frac{1}{2}(\nabla_\gamma\Phi\nabla^\gamma\Phi+m^2 \Phi^2)
\]
one notes that it contains terms quadratic in $\Phi(x)$ and its
derivatives, but $\la\psi|\hat{\Phi}(x)\hat{\Phi}(x)|\psi\ra$ is in
general not defined (remember that
$\la\psi|\hat{\Phi}(x)\hat{\Phi}(y)|\psi\ra$ is a
distribution). Therefore, it was an old idea to admit only those
states $\psi$ as physical whose two-point functions $\Lambda^{(2)}$
have all the same singular kernel $G(x,y)$ and differ only in their
smooth parts, since then one can unambiguously define
\[ \la\psi|:\hat{\Phi}(x)\hat{\Phi}(x):|\psi\ra:=\lim_{x\to
  y}\left[\la\psi|\hat{\Phi} (x)\hat{\Phi}(y)|\psi\ra-G(x,y)\right]
\]
and $\la\psi|:\hat{T}_{\mu\nu}(x):|\psi\ra$ on the r.h.s.~of \rf{3.1}
can be made a well defined quantity. These are the so-called {\it
  Hadamard states}. Without giving their precise definition in terms
of their singular kernel (which can be found in \cite{KW91}) we state
the following important theorem which was recently proven by
M. Radzikowski \cite{Rad96} and which characterizes these states in a
local and covariant manner by
the wavefront set of $\Lambda^{(2)}$:
\begin{theorem}[Theorem 5.1 of \cite{Rad96}]\\
A quasifree state of the linear Klein-Gordon quantum field on a
globally hyperbolic spacetime manifold $(\M,g)$ is an Hadamard state
\beq
 \Leftrightarrow WF(\Lambda^{(2)})=\left\{(x_1,k_1;x_2,-k_2)\in
T^*(\M\times\M)\setminus\{0\}; (x_1,k_1)\sim (x_2,k_2);k_1^0\geq
0\right\}\label{3.2} 
\eeq
where $(x_1,k_1)\sim (x_2,k_2) :\Leftrightarrow x_1$ and $x_2$ are
connected by a null geodesic $\gamma$, $k_1, k_2$ are cotangent to
$\gamma$ at $x_1$ resp.~$x_2$ and parallel transported to each other
along $\gamma$.
\end{theorem}
(Note that a globally hyperbolic manifold is time orientable, hence
the condition $k_1^0\geq 0$ makes good sense.)
This theorem says that only {\it positive} frequencies occur in
$WF(\Lambda^{(2)})$. This is the sought for microlocal remnant of the
spectrum condition.

\newsection{Construction of Hadamard states}
Having recognized the physical importance of Hadamard states we are
immediately led to the following mathematical problem: Construct (if
possible all) Hadamard states for a given spacetime $(\M,g)$,
i.e.~(all) $\Lambda^{(2)} \in \Dp{\M\times\M}$ fulfilling the
following five conditions (which are the axioms (i)--(v) from above
rewritten in terms of $\Lambda^{(2)}$):\\
(i) $\Lambda^{(2)}(f_1,f_2)=\ol{\Lambda^{(2)}(f_2,f_1)}\quad \forall
f_1,f_2\in \D{\M}$ (hermiticity)\\
(ii) $\lt(f,f)\geq 0\quad \forall f\in \D{\M}$ (positivity of scalar
product in ${\cal H}$)\\
(iii) Im$\lt(f_1,f_2)=\frac{1}{2}\la f_1, E f_2\ra$ (commutation relations)\\
(iv) $\lt((\Box_g+m^2)f_1,f_2)=0=\lt(f_1,(\Box_g+m^2)f_2)$ (field
equations)\\
(v) $WF(\lt)$ as in Eq. \rf{3.2} (microlocal spectrum condition).\\  \\
To solve this problem we first give a parametrization of (pure) 
quasifree states in terms of two operators $R$ and $I$:
\begin{theorem}[Theorem 3.11 of \cite{Junker96}]\\ \label{Theorem4.1}
Let $(\M,g)$ be a globally hyperbolic spacetime with Cauchy surface
$\Sigma$ (and unit normal $n^\mu$ on $\Sigma$). Let $R$ be a
selfadjoint and $I$ a selfadjoint, positive and invertible operator on
$L_{\R}^2(\Sigma,d^3\sigma)$ ($d^3\sigma$ the volume measure on
$\Sigma$ induced by $g$). Then
\beq
\ls(f_1,f_2)=\frac{1}{2}\left\la(R-iI-n^\mu \nabla_\mu)Ef_1,
I^{-1}(R-iI-n^\mu\nabla_\mu)
Ef_2\right\ra_{L^2_{\C}(\Sigma,d^3\sigma)} \label{5.1}
\eeq
is the two-point function of a pure quasifree state, i.e.~fulfills
(i)--(iv). 
\end{theorem}
A proof of this theorem and the following ones and a generalization to
mixed states can be found in \cite{Junker96}. Next we give conditions
on $R$ and $I$ such that $\ls$ of Eq. \rf{5.1} has the correct wavefront set
(property (v) above). The idea is to project out from the fundamental
solution $E$ (whose WF is known to contain positive {\it and} negative
frequencies) only the positive frequencies.
 To this end we consider a foliation of $\M$
into hypersurfaces $\Sigma_t$, $t\in ]-T,T[\subset\R$, in a small
neighborhood of the Cauchy surface $\Sigma$: $\M=]-T,T[\times\Sigma,\,
\Sigma_t=\{t\}\times\Sigma$, and take the operators $R$ and $I$ as
depending on the parameter $t$.

\begin{theorem}[Theorem 3.12 of \cite{Junker96}] \\
\label{Theorem4.2}
Let $I(t), R(t)$ be pseudodifferential operators on $\Sigma_t$,
$t\in]-T,T[$ such that $I$ is elliptic and such that there exists a
pseudodifferential operator $Q$ on $]-T,T[\times\Sigma$ which has the
property
\beq
Q(R-iI-n^\mu\nabla_\mu) =\Box_g+m^2+r \label{5.2}
\eeq
for some smoothing operator $r$, and which possesses a principal symbol
$q(x,k)$ with the property
\beq
q^{-1}(0) \setminus\{0\}\subset \{(x,k)\in T^*\M;k^0>0\}. \label{5.3}
\eeq
Then $\ls$, Eq. \rf{5.1}, is an Hadamard state (i.e.~$\ls$ has the
wavefront set \rf{3.2}).
\end{theorem}

The sufficient conditions of this theorem allow to prove the Hadamard
property for many examples of quantum states which have already been
constructed in the literature on certain spacetime manifolds (in the
sense of Theorem \ref{Theorem4.1}),
but which have so far not been shown to be of the Hadamard type (which is a
necessary condition for being 
physically acceptable). Among these are the ground and thermodynamic
equilibrium states on static spacetimes \cite[Theorems 3.18 and 3.19]
{Junker96} 
and the so-called adiabatic
vacuum states on Robertson-Walker spacetime models \cite[Theorem 3.24]
{Junker96}. It can also be shown that the frequently employed method
of Hamiltonian diagonalization for constructing quantum states on a
curved manifold does in general not lead to Hadamard states and is
therefore unphysical \cite[Theorem 3.27]{Junker96}.\\
But even more important, Theorem \ref{Theorem4.2} gives us a guide to a 
method of explicitly constructing Hadamard states on an arbitrary globally
hyperbolic spacetime manifold. The idea is to construct the
pseudodifferential operators
$R(t)$, $I(t)$ and $Q$ by an asymptotic expansion of their symbols
such that Eq.~\rf{5.2} is fulfilled.
To achieve this we choose Gau{\ss}ian normal coordinates in a
neighborhood of $\Sigma$ such that the metric reduces to the simple
form
\[ g_{\mu\nu} = \left(\begin{array}{cc}1& \\ & -h_{ij}(t,\vec{x})
\end{array}\right) \]
where $h_{ij}(t,\vec{x})$ is the Riemannian metric on $\Sigma_t$
induced by $g$, and the Klein-Gordon operator reads
\beq
\Box_g+m^2 
= \frac{1}{\sqrt{h}}\partial_t(\sqrt{h}\partial_t \cdot)-\frac{1}{\sqrt{h}}
\partial_i(\sqrt{h} h^{ij}\partial_j \cdot)+m^2.\label{5.4}
\eeq
Then we make the following Ansatz for a factorization of \rf{5.4}
\beqa
\Box_g+m^2 &=& (-a-\frac{1}{\sqrt{h}}\partial_t \sqrt{h})
(a-\partial_t)-r \label{5.5}\\
a(t,x,k) &\sim& -i\sqrt{h^{ij}k_i k_j +m^2} +\sum_{\nu=1}^\infty
b^{(\nu)}(t,x,k) \nonumber
\eeqa
with $r$ a smoothing operator and
 $b^{(\nu)}$ symbols of order $1-\nu$. It turns out that the
 $b^{(\nu)}$ can be determined successively such that Eq.~\rf{5.5}
 holds, and that the operators $I, R, Q$ corresponding to the symbols
\beqa
I(t,x,k) &:=& -\frac{1}{2i}\left[a(t,x,k)-\ol{a(t,x,-k)}\right]\nonumber\\
R(t,x,k) &:=& \frac{1}{2} \left[ a(t,x,k) + \ol{a(t,x,-k)}\right] \label{5.6}\\
Q(t,x,k) &:=& -a(t,x,k)-\frac{1}{\sqrt{h}}\partial_t\sqrt{h}\nonumber
\eeqa
satisfy all the requirements of Theorems \rf{Theorem4.1} and
\rf{Theorem4.2} (for details see \cite[Section 3.7]{Junker96}).
Therefore we have the following
\begin{theorem}[Theorem 3.29 of \cite{Junker96}]\\ \label{Theorem4.3}
Let $(\M,g)$ be a globally hyperbolic spacetime with Cauchy surface $\Sigma$.
In a neighborhood of $\Sigma$ let the pseudodifferential operators $I$
and $R$ be constructed as shown in Eq.~\rf{5.6}.\\
Then Eq.~\rf{5.1} is the two-point distribution of a pure Hadamard
state of the Klein-Gordon quantum field on $(\M,g)$.
\end{theorem}

\newsection{Outlook}
The new characterization of physical quantum states in gravitational
background fields by the wavefront set opens a new era in the study
of QFT on curved spacetimes. The following problems can and will be
treated on globally hyperbolic manifolds using microlocal techniques:\\
- generalization to self-interacting scalar quantum fields
(e.g.~$\Phi^4$-theory) by giving a microlocal spectrum condition for
all $n$-point functions \cite{BFK96}\\
- perturbation and renormalization theory (R.~Brunetti \&
K.~Fredenhagen, in preparation)\\
- generalization of the results of Section 4 to Dirac- and
electromagnetic fields on manifolds using the polarization set\\
- construction of states on physically interesting spacetime models and
calculation of physical effects\\
- construction of a Euclidean version of QFT on curved spacetimes
(W. Junker, in preparation)\\
- treatment of (non-abelian) gauge-theories in this frame...



\noindent{\bf Addresses:}

{\sc Wolfgang Junker,} Max-Planck-Institut f\"ur Gravitationsphysik,
Albert-Einstein-Institut, Schlaatzweg 1, D-14473 Potsdam, Germany,
e-mail: junker@aei-potsdam.mpg.de


\end{document}